\documentclass[sigconf, nonacm]{acmart}
\usepackage{algorithm}
\usepackage{algpseudocode}
\usepackage[export]{adjustbox}

\usepackage{graphicx}
\usepackage{subcaption}

\newcommand{\sunny}[1]{[\textcolor{teal}{#1}]}

\usepackage{tabularx,booktabs}
\AtBeginDocument{%
  \providecommand\BibTeX{{%
    Bib\TeX}}}

\setcopyright{acmlicensed}
\copyrightyear{2018}
\acmYear{2018}
\acmDOI{XXXXXXX.XXXXXXX}
\acmConference[Conference acronym 'XX]{CIDR}{June 03--05,
  2018}{Woodstock, NY}
\acmISBN{978-1-4503-XXXX-X/2018/06}

\begin{document}

\title{Data-Aware Socratic Query Refinement in Database Systems}

\author{Ruiyuan Zhang}

\affiliation{%
  \institution{Hong Kong Generative AI Research and Development Center }
  \city{Hong Kong SAR}
  \country{China}
}
\email{zry@hkgai.org}

\author{Chrysanthi Kosyfaki}
\affiliation{%
  \institution{Hong Kong University of Science and Technology}
  \city{Hong Kong SAR}
  \country{China}}
\email{ckosyfaki@cse.ust.hk}

\author{Xiaofang Zhou}
\affiliation{%
  \institution{Hong Kong University of Science and Technology}
  \city{Hong Kong SAR}
  \country{China}}
\email{zxf@cse.ust.hk}

\renewcommand{\shortauthors}{Zhang et al.}

\begin{abstract}
In this paper, we propose Data-Aware Socratic Guidance (DASG), a dialogue-based query enhancement framework that embeds \linebreak interactive clarification as a first-class operator within database systems to resolve ambiguity in natural language queries. DASG treats dialogue as an optimization decision, asking clarifying questions only when the expected execution cost reduction exceeds the interaction overhead. The system quantifies ambiguity through linguistic fuzziness, schema grounding confidence, and projected costs across relational and vector backends. Our algorithm selects the optimal clarifications by combining semantic relevance, catalog-based information gain, and potential cost reduction. We evaluate our proposed framework on three datasets. The results show that DASG demonstrates improved query precision while maintaining efficiency, establishing a cooperative analytics paradigm where systems actively participate in query formulation rather than passively translating user requests.
\end{abstract}



\maketitle

\section{Introduction}
Data engineers increasingly rely on large language models (LLMs) for routine query construction, achieving significant productivity improvements. Modern LLM-based platforms enable analysts to submit natural-language questions directly to data warehouses and obtain executable SQL through single-prompt interactions \cite{Li2024}\cite{Gao2024}. Recent advances in language models, including GPT-4o \cite{openai2024gpt4technicalreport}, Gemini 2.5 Pro \cite{comanici2025gemini25pushingfrontier}, Qwen2.5 Coder \cite{hui2024qwen2}, and retrieval-augmented decoding, have driven the accuracy of the exact-match beyond 90\% on Spider \cite{yu-etal-2018-spider}, nearly 80\% on BIRD \cite{li2024can}, and nearly 60\% on Spider 2 \cite{lei2024spider} benchmarks. 

Although the "guessing ability" of LLMs continues to improve, enabling more effective translation of natural language into SQL, non-expert users still frequently face a frustrating cycle of repeated attempts and failures. This challenge does not stem from syntactic errors in the generated SQL, but rather from a fundamental misalignment between users' conceptual understanding of data and the precise demands of database execution engines. Without familiarity with underlying database schemas, SQL semantics, or effective prompting techniques, users often formulate uncertain queries that yield incomplete or incorrect results. These ambiguous inputs lead to unguided iterations that rarely converge on satisfactory solutions.

Consider a scenario in which a non-database-expert user submits the query: ``Show me active premium users for the last quarter in Germany.'' The user has only a general understanding of the database’s topic but no knowledge of its underlying schema. As a result, they are unable to provide the precise instructions needed for an LLM to generate an accurate SQL query such as:  
"Pulls the \texttt{user\_ids} column from the \texttt{customers} table for every row with \texttt{country\_code = 'DE'}, \texttt{plan\_type = 'premium'}, and \texttt{last\_login} between \texttt{2025-04-01} and \texttt{2025-06-30}". Similarly, an uncertain request like \textit{``Show me some unique customers''} presents significant interpretive challenges: does ``unique'' refer to de-duplicated customer identities, one-time purchasers, or behaviorally anomalous users?

This ambiguity creates several key challenges. First, it hinders large language models from generating accurate and executable SQL, as the gap between natural language and formal query logic remains unaddressed. Second, beyond correctness, ambiguous queries pose serious operational risks. Even if the generated SQL is valid, vague intent can lead to expensive full-table scans or unbounded data retrieval, consuming excessive resources and potentially violating performance service-level agreements. As a result, ambiguity harms not only accuracy but also system efficiency and scalability. Third, while some prior work has explored ambiguity in queries, most approaches assume a traditional relational database. However, modern AI systems increasingly rely on vector databases for semantic search over unstructured data like text, images, and audio. In this context, ambiguity manifests differently—through mismatched embeddings or unclear cluster boundaries—leading to irrelevant or unstable results. So far, little attention has been paid to how such ambiguity should be handled in vector search. To ensure efficient and accurate retrieval, systems must actively detect and resolve ambiguity through interactive, data-driven clarification. This highlights the need for a unified, multi-modal approach to query understanding across both structured and semantic data systems.

Therefore, we identify the following key limitations in current NL2SQL related systems: \textbf{Single-pass assumption}: Systems proceed to execution immediately after the first parsing attempt, ignoring the inherently iterative nature of real-world data exploration. \textbf{Language-centric uncertainty}: Follow-up questions are typically derived from model logits (e.g., LLM confidence scores) rather than data-driven statistics, leading to clarifications that refine wording without reducing the actual answer space. \textbf{Backend rigidity}: Most solutions are designed exclusively for relational databases, despite the growing prevalence of hybrid architectures that integrate tables with vector indexes for multimodal data.

In this paper, we integrate ambiguity resolution directly into the database system, treating clarification as a first-class physical operator. Guided by the principle that systems—not users—should ensure query precision, we introduce Data-Aware Socratic Guidance (DASG), an interactive framework that embeds follow-up questions as native optimization decisions in query execution. DASG addresses three key gaps unmet by current state-of-the-art approaches:

\begin{itemize}
  \item \textbf{DASG Framework.} We introduce \emph{Data-Aware Socratic Guidance (DASG)}, the first system that embeds a \emph{Conversational Query Optimizer (CQO)}—deciding whether to ask the user for clarification—and a \emph{Maximum Information Utility (MIU) Ranker}—selecting the best clarification—directly into the DBMS query pipeline.
  \item \textbf{Cost-driven clarification planning.} CQO applies an cost model that compares the expected benefit of a clarification against its interaction cost, using catalog statistics and cost estimates to make principled decisions about when to engage the user.
  \item \textbf{Data-grounded ambiguity control and hybrid querying.} DASG quantifies query uncertainty using database histograms, learned cardinalities, and vector densities to ensure each clarification meaningfully reduces result \linebreak ambiguity and generates predicates that natively target both SQL filters and vector index operations.
\end{itemize}

The paper is organized as follows. 
Section~\ref{sec:related} reviews related work in interactive NL2SQL and AI-enhanced DBMSs. 
Section~\ref{sec:paradigm} introduces the Co-operative Analytics paradigm. 
Section~\ref{sec:framework} presents the DASG framework, including its dialogue manager, cost model, and hybrid backend integration. 
Section~\ref{sec:evaluation} reports preliminary results on four benchmarks. 
Section~\ref{sec:conclusion} concludes and outlines future work.

\section{Related work}
\label{sec:related}

Recent advances in AI have made LLM-based interfaces a popular choice for database access. As language models improve in code generation, their use in Text-to-SQL tasks has surged. Execution accuracy has risen from ~60\% to over 80\% on cross-domain benchmarks like Spider and DAIL-SQL \cite{gao2023texttosqlempoweredlargelanguage}. Liu et al.\ \cite{Liu2020} survey over fifty LLM-to-SQL methods, including prompt-only agents, instruction-tuned models, and hybrid symbolic–neural pipelines. Still, even the strongest systems fail when faced with ambiguous phrasing, missing schema context, or expensive query plans—issues highlighted by the BIRD benchmark, which penalizes inefficient SQL generation.

Recent text-to-SQL work has explored interactive frameworks that incorporate user feedback. Early systems allowed correction via error tagging \cite{iyer2017learningneuralsemanticparser} or structured edits \cite{Li2014}. Later methods used multiple-choice prompts to resolve ambiguity \cite{yao-etal-2019-model,gur-etal-2018-dialsql}, but limited feedback modes restrict their ability to handle complex errors. NL-EDIT \cite{elgohary-etal-2021-nl} accepts free-form revisions, offering more flexibility—but interpreting open-ended feedback remains as hard as the original query. In contrast, our approach targets a middle ground: asking precise, data-driven clarification questions while keeping user input minimal.
Recent work has advanced interactive text-to-SQL systems. Tian et al. \cite{tian2024interactivetexttosqlgenerationeditable} proposed STEPS, which decomposes SQL generation into interpretable, editable steps aligned with SQL clauses. Qiu et al. \cite{qiu2025interactivetexttosqlexpectedinformation} introduced a probabilistic method that uses expected information gain to select optimal clarification questions, enabling efficient, transparent, and globally informed dialogues.

A parallel trend integrates AI directly into the DBMS. "LLM-as-table" treats models as virtual tables that can be joined via prompts. NeurDB \cite{neurdb-scis-24} envisions a learned query optimizer for autonomous databases. Commercial systems like Microsoft Fabric Copilot, Snowflake Cortex, and Google BigQuery’s AI interface now support natural language analytics. Meanwhile, TableGPT \cite{tablegpt} and related efforts \cite{Lu2025} fine-tune LLMs on synthetic table tasks to improve understanding of structured data and boost performance on table-based reasoning.

\section{The Cooperative Analytics Paradigm}
\label{sec:paradigm}

As mentioned in Section \ref{sec:related}, most prior work treats the LLM as an external assistant: the model emits SQL, the engine executes, and the interaction ends. In this workflow, any ambiguity in the user’s intent remains invisible to the database engine, and the user must rephrase or issue a new query entirely on their own. Such one-shot designs limit both the precision of the results and the efficiency of the system.

We advocate a different approach: one that closes the loop between user, model, and engine. Instead of viewing the LLM as a “black box” query generator, we place it as the first partner in a cooperative cycle. The engine inspects the generated query, leverages catalog statistics or vector-index summaries to measure uncertainty, and—only when needed—asks the user a short, targeted question. This light interaction reduces wasted execution time and guides users toward precise results without requiring them to frame perfect queries up front.

\begin{figure*}[htbp]
  \centering
  \includegraphics[width=0.9\textwidth]{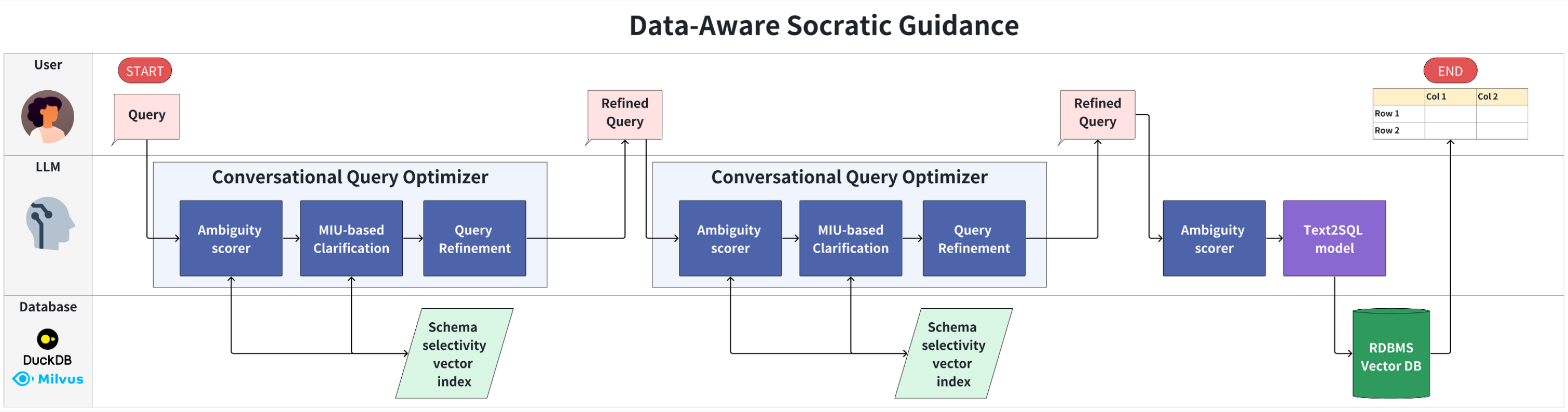}
  \caption{Data-Aware Socratic Guidance Framework}
  \label{fig:framework}
\end{figure*}

These developments suggest a shift from pure automation toward human-AI collaboration. We propose a Cooperative Analytics Paradigm in which human users, large language models, and the database engine form a closed-loop interaction: the model interprets user intent, the engine identifies and diagnoses uncertainty, and the user provides lightweight clarification. By integrating this cycle into the query processing pipeline, users benefit from a smoother and more intuitive experience, no longer required to formulate fully precise or technically detailed queries. At the same time, the system avoids redundant computations, unnecessary query executions, and endless rounds of clarification that typically arise from ambiguous inputs.

Together, these trends point to a Cooperative Analytics Paradigm grounded in four key principles:

\begin{enumerate}
\item \textbf{Shared responsibility for precision}
Database systems must actively detect underspecified user intent and initiate the refinement process. Users articulate high-level goals, while the system determines what additional information is needed to proceed effectively.
\item \textbf{Data-grounded guidance}
Clarification requests should be informed by analysis of catalog statistics or summaries from vector indexes, rather than relying solely on language model heuristics. The central criterion is \emph{result-set entropy}: if many tuples remain plausible answers, the query is still ambiguous and further clarification is warranted.
\item \textbf{Cost awareness}
Each interaction in the dialogue—each clarification turn—incurs real costs, including LLM token usage, network latency, and user attention. These can be quantified using the same cost model as traditional database operations such as I/O and CPU. The system should issue a clarification request only when the expected reduction in overall execution cost exceeds the overhead of the interaction.
\item \textbf{Backend agnosticism}
Modern applications often combine structured data with embedding spaces. A robust refinement mechanism must therefore generate appropriate conditions for both SQL predicates over relational tables and similarity filters over vector databases, ensuring compatibility across heterogeneous backends.
\end{enumerate}

\section{DASG Framework Architecture}
\label{sec:framework}
To put the Cooperative Analytics Paradigm into practice, we build Data-Aware Socratic Guidance (DASG), shown in Figure \ref{fig:framework}. DASG works with an existing SQL engine (we use DuckDB in our experiments) and a vector database, coordinating five components in a closed loop. The system helps users refine vague queries through smart, data-driven questions.

\subsection{The Conversational Query Optimizer}

CQO intercepts the query before execution and decides whether to ask exactly one follow-up question. It predicts the cost of running the original plan using catalogue statistics and a language-model ambiguity score; if the plan appears cheap or precise, CQO does nothing. When the plan looks both expensive and under-specified, CQO selects the single facet—department, time interval, margin band, or similar—that its ranking model expects will cut the result space most per unit of compute. The user answers, the new predicate is injected into SQL or ANN filters, and the refined plan runs once.

Clarification is triggered when the Value of Clarification (VoC) exceeds the Cost of Dialogue (CoD). VoC includes latency saved, quality gain, and reduced user effort: $ w_h \log N $, where $ N $ is the number of retries a user would make and $ w_h $ weights cognitive cost. The log term captures diminishing returns---avoiding early retries matters most. CoD accounts for conversation time (priced at user time cost) and the small LLM call expense. CQO asks questions only when this trade-off is positive, i.e. $VoC > CoD$, making clarification an optimization, not a patch---ensuring users are never slowed or burdened. The model adapts to context: high-stakes apps (e.g., BI) use higher $ w_h $ and clarify early; low-risk ones (e.g., e-commerce) intervene only if costly operations loom. Simple, unambiguous queries require no dialogue. Thus, CQO extends cost-based planning into interaction, asking questions only when they yield faster or better outcomes.In this preliminary study, we focus on latency as the primary cost; human effort, energy, and computational costs will be incorporated in future work.

\subsection{Ambiguity Evaluator}

Before deciding whether to ask for more information, DASG first measures how unclear the user’s request is. It checks three sources of ambiguity. Ambiguity is treated as a combination of three signals: linguistic vagueness, schema-grounding confidence, and the database’s own cost predictions. Any one of these signals, if high enough, is sufficient to trigger a follow-up question. 

This work extends beyond RDBMS to vector databases, where queries typically involve semantic similarity or nearest-neighbor searches. In this context, ambiguity arises from semantic ambiguity, where the user’s intent, encoded as an embedding, may only loosely match stored vectors due to subtle differences in meaning, and distribution ambiguity, where queries fall in dense regions or between clusters, making intent unclear. These uncertainties degrade retrieval relevance and efficiency. Like relational queries, effective vector search requires an iterative clarification process—guided by data-aware metrics—to refine intent and improve both accuracy and performance.

\subsubsection{Linguistic ambiguity}
The first signal is independent of any particular database. We feed the raw user sentence to a large language model that has been prompted to behave as a “vagueness scorer.” The model returns a scalar between 0 (clear and specific) and 1 (highly underspecified). Prompts focus the model on well-known linguistic cues—indefinite adjectives (“some,” “recent”), missing comparators (“large orders”), or nested references without clear scope (“top customers with high value”). This score flags queries that are ambiguous even in ordinary language.

\subsubsection{Schema-grounding confidence}

Next we ask how well the terms in the request line up with objects in the catalogue. A lightweight entity linker extracts each noun phrase and salient adjective, then attempts to map the phrase uniquely to a table name, a column name, or a plausible value inside a column. The schema-grounding score is the fraction of key terms anchored in this way. A low score indicates that the user’s mental model of the data separates from the actual schema, an early warning that structural guidance will be required.

\subsubsection{Estimated execution cost}

DASG next estimates the \emph{operational} risk of executing the draft query as‐is—\textbf{for both relational and vector operators}.

\begin{itemize}
  \item \textbf{Relational path.}  
        A baseline Text-to-SQL model emits a provisional statement.  
        Instead of running it, the system issues  
        \texttt{EXPLAIN (FORMAT JSON)} and extracts  
        (i) the root cardinality(expected result rows), histogram, complexity etc,   
        (ii) the total estimated cost, and
        (iii) the highest cost among any single operation.
        The third value flags poor join orders that a simple total cost may hide.
  \item \textbf{Vector path.}  
        When the query uses an embedding column, DASG checks the ANN index (e.g., HNSW or IVF-FLAT) for:flat) and retrieves  
        (i) expected number of index probes or hops on the graph,
        (ii) average number of candidate results
\end{itemize}

These three signals—language clarity, schema match, and execution cost—are combined using a weighted sum as the \textbf{ambiguity score} of user's query. 

As discussed in this paper, we extend the concept of ambiguous queries beyond relational databases to vector databases. In vector databases, users typically issue queries such as semantic similarity searches or nearest-neighbor (NN) lookups. Ambiguity in this context arises from two sources: \textbf{Semantic Ambiguity}, where a user's intent—encoded as a vector embedding via an LLM or embedding model—may only loosely align with stored embeddings due to subtle meaning differences; and \textbf{Distribution Ambiguity}, where queries fall in high-density regions or near boundaries between semantic clusters, making the intended results uncertain. These types of ambiguity directly impact retrieval efficiency and relevance. Therefore, just as relational queries benefit from systematic ambiguity reduction, vector search queries also require an iterative, clarity-driven process guided by metrics tailored specifically for vector data.

\subsection{The Maximum Information Utility (MIU) Ranker}
Once CQO’s economic gate confirms that asking is worthwhile, control passes to the Model of Intent Uncertainty (MIU).  MIU’s task is to pick \emph{which} follow-up question will squeeze the most work out of that single dialogue turn.  A candidate question always targets one \emph{facet}—a column in a SQL table, an attribute group such as “order timing,” or a numeric filter on a vector index—that can refine the current predicate set.  Choosing the right facet is critical: ask the user about the wrong dimension and the plan stays slow; ask about the right one and the result space collapses.

MIU therefore constructs a set \(\mathcal{F}=\{f_1,\dots,f_k\}\) of candidate facets still under-specified by the user’s wording.  For each \(f_i\) it computes a three-part score.  First, \textbf{semantic alignment} measures how well the facet’s name or description matches the query embedding, rewarding questions the user is likely to understand instantly.  Second, \textbf{expected information gain} estimates how much the result set would shrink if the facet were pinned down; the estimator uses catalogue histograms for relational data and posting-list fan-outs for vectors.  Third, \textbf{filter cost} predicts the extra runtime of applying the facet as a predicate.  The combined score is  

\[
S(f_i)=\alpha\,\text{align}(f_i)
        +\beta\,\text{gain}(f_i)
        -\gamma\,\text{cost}(f_i),
\]

where \(\alpha,\beta,\gamma\) are weights learned offline from past queries.

Because different facets differ wildly in units, MIU normalises each term, then selects the single facet \(f^\star=\arg\max_{f_i\in\mathcal{F}} S(f_i)\). 

Although inspired by relational query optimizers, MIU's scoring framework operates directly on database statistics rather than abstract, backend-agnostic hooks. It tightly integrates with DuckDB to collect real-time cardinality, selectivity, and distribution data, then computes a score using a weighted model. Instead of relying on generic estimation interfaces, MIU analyzes DuckDB's \texttt{EXPLAIN} plans, column statistics, and Gini coefficients to derive precise cost and selectivity estimates.

MIU returns the top-ranked clarification facet along with a normalized score $m \in [0,1]$. The Conversational Query Optimizer (CQO) uses this score as a confidence threshold: if $m < 0.5$, the system skips clarification, avoiding speculative or low-value interactions. In this role, MIU acts as the intent uncertainty quantifier in the conversational stack, enabling CQO to allocate user attention only when the expected value of clarification outweighs its cost.

\subsection{The Question Synthesizer}
The Question Synthesizer turns the facet selected by the MIU ranker into a clear, single-sentence prompt for the user. It receives two inputs: (i) a small structure that names the chosen clarification target and any associated filter candidates, and (ii) the running conversation history. These are inserted into a fixed prompt template and passed to a large language model, which returns a concise, human-readable question. For example, if the target facet is the column Gross, the Synthesizer might ask:
“To clarify what you mean by ‘successful,’ should we focus on the financial gross?”

By delegating surface‐form generation to the language model, this component cleanly separates the analytical choice of what to ask from the linguistic task of how to ask it.

\section{Evaluation}
\label{sec:evaluation}

We tested DASG using three publicly available datasets that collectively cover relational single-table operations, multi-table queries, and vector-based tasks. For each dataset, we constructed a natural-language query workload and assessed performance across three key metrics: latency, answer quality, and dialogue cost.

\subsection{Datasets}

\begin{table}[t]
\small
\setlength{\tabcolsep}{4pt}
\caption{Datasets used in our experiments}
\label{tab:datasets}
\begin{tabular*}{\columnwidth}{@{\extracolsep{\fill}} l r l l}
\toprule
\textbf{Dataset}   & \textbf{Size} & \textbf{Category} & \textbf{Type} \\ \midrule
IMDb--TMDb *10     & 11\,M rows    & Movies     & Relational \\
Instacart          & 3.2\,M orders  & Commerce   & Relational \\
SciDocs (BEIR)     & 25.7\,k docs   & Scholarly  & Vector     \\
\bottomrule
\end{tabular*}
\vspace{-0.8em}
\end{table}

We evaluated our approach on three diverse real-world datasets: IMDb, Instacart, and SciDocs. The IMDb dataset is a 10× expanded version with 85,855 movies and 70 test queries, providing a rich schema for complex relational queries. The Instacart dataset contains 3.4 million grocery orders and 1.4 million products, with 50 test queries that reflect realistic user intents in a large-scale e-commerce setting. SciDocs includes 24,657 scientific papers with 768-dimensional embeddings, paired with 50 naturally vague queries and their refined versions, enabling evaluation of ambiguity resolution in semantic search over vector spaces. 

\subsection{Workload Generation}

We built three query sets. IMDb contains 75 hand-written questions that mix vague terms, missing filters, and implicit conditions; one-third are purposely high-cost. Instacart adds 50 questions following the same ambiguity pattern but engineered to trigger multi-table joins. SciDocs contributes 50 queries automatically generated from BEIR document pairs; we inject mild topic and terminology fuzziness and keep brute-force results as ground truth.

\subsection{Metrics}
We investigated system performance using two primary metrics: \textbf{latency} and \textbf{recall@100}. Latency measures the end-to-end response time, including ambiguity detection, clarification (if triggered), and query execution, reflecting the system’s interactivity and efficiency. Recall@100 is used to assess result quality, defined as the fraction of ground-truth relevant items that appear within the top 100 returned results. 

\subsection{Implementation Details}
We used DuckDB for relational queries and FAISS (with IVF-Flat and HNSW indices) for vector search. Experiments run on a server with 2× Intel Xeon Gold 6342 CPUs, 512 GB RAM, and Qwen3-32B (via API) for language processing.

The MIU Ranker uses weights: $\alpha=0.4$ (relevance), $\beta=0.4$ (information gain), and $\gamma=0.2$ (execution cost). The Ambiguity Evaluator uses $\delta=0.4$ (linguistic ambiguity), $\varepsilon=0.4$ (schema grounding), and $\zeta=0.2$ (projected cost). These weights, determined empirically on a validation set, balance clarity, effectiveness, and efficiency in the clarification process. For simplicity and focus, this work considers only latency as the main cost factor. Human interaction costs and other overheads—such as energy and computational expense—are left for future extension. 

\subsection{Preliminary Results}

\subsubsection{IMDB dataset}

\begin{figure}[t]
  \centering
  \includegraphics[width=0.8\linewidth]{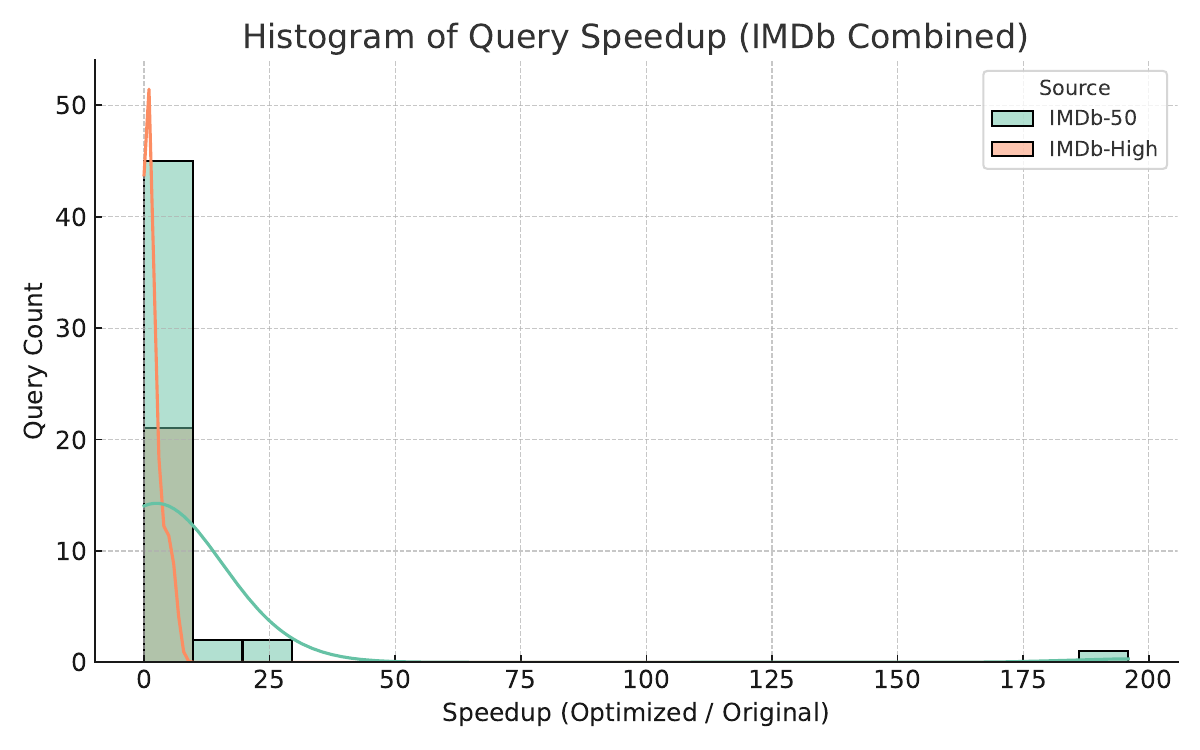}
  \caption{Histogram of Query Speedup --- IMDB}
  \label{fig:imdb_total_1}
\end{figure}

Figure~\ref{fig:imdb_total_1} shows a right-skewed speed-up distribution: most queries gain 1.2×–2×, with a long tail reaching up to 190×. CQO delivers consistent improvements across query costs, with minor regressions only in low-latency cases, where dialogue overhead is relatively high. The largest gains occur in moderate- to high-cost queries, where one clarification avoids costly, inefficient operations. This demonstrates the effectiveness of the single-clarification strategy and validates CQO’s core idea: well-placed questions can drastically reduce the plan space and yield significant performance gains. 

\begin{figure}[t]
  \centering
  \includegraphics[width=0.8\linewidth]{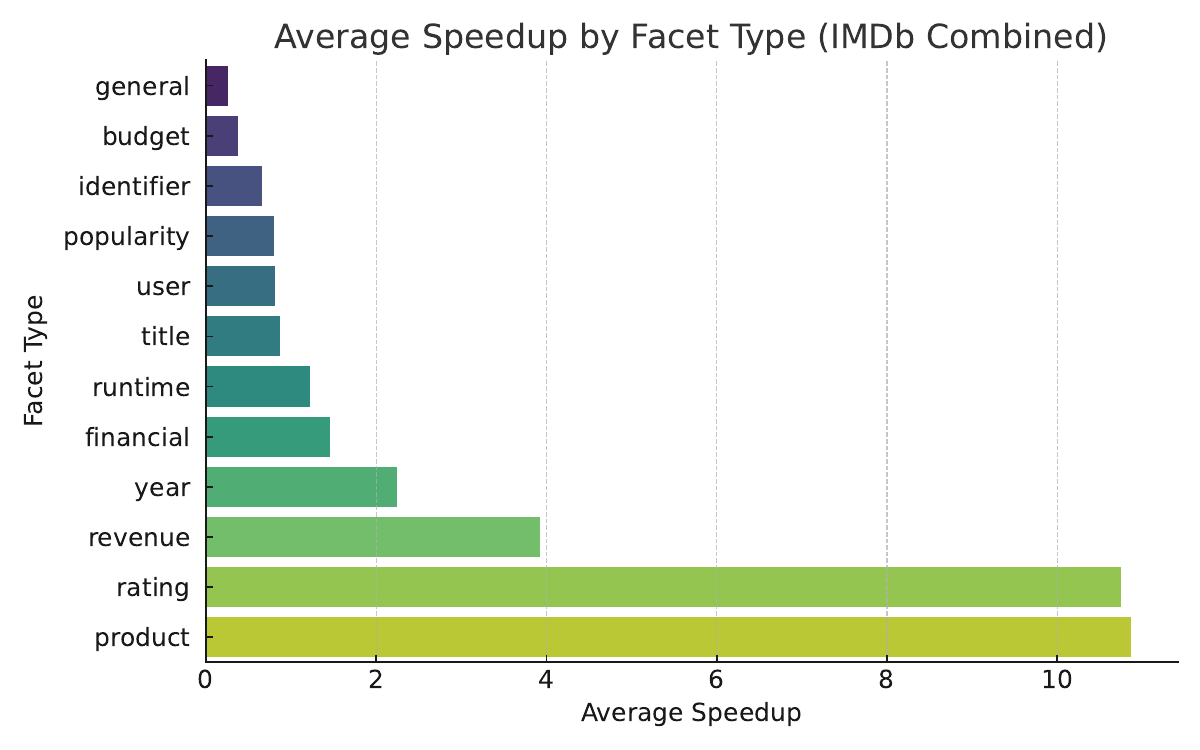}
  \caption{Average Speedup by Facet Type --- IMDB}
  \label{fig:imdb_total_2}
\end{figure}

Figure~\ref{fig:imdb_total_2} shows significant variation in average speed-up across facet types. Attributes like \texttt{rating} and \texttt{runtime} achieve high gains (often >2.0×), as they filter unindexed, low-selectivity columns and reduce expensive operations. In contrast, facets like title—already indexed and selective—yield little benefit or cause regressions. This highlights the importance of facet semantics in optimization effectiveness and suggests that the MIU ranker can be improved by more strongly weighting attributes with higher execution impact.

\subsubsection{Instacart dataset}

\begin{figure}[t]
  \centering
  \includegraphics[width=0.82\linewidth]{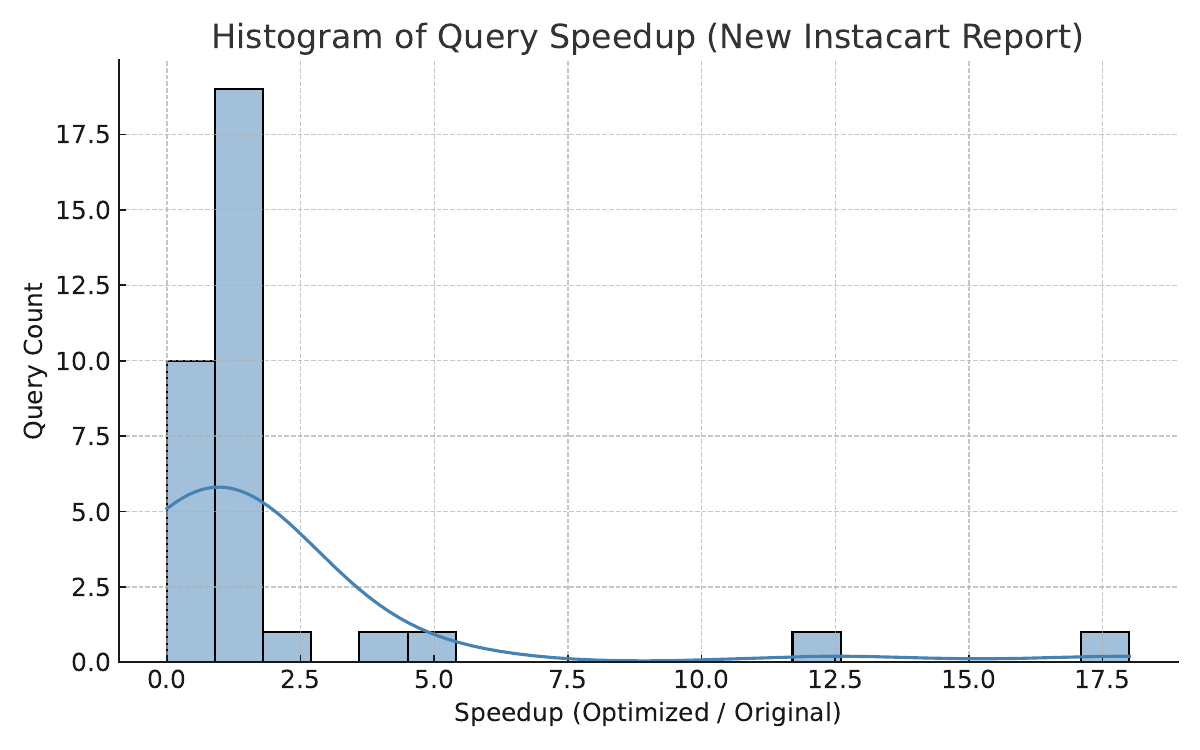}
  \caption{Histogram of Query Speedup --- Instacart}
  \label{fig:Instacart_1}
\end{figure}

Figure~\ref{fig:Instacart_1} shows the distribution of query speedups on the Instacart workload. The histogram is centered slightly above 1.0×, with most queries gaining between 1.2× and 1.6×, a few exceeding 2×, and a left tail indicating mild regressions (0.8×–1.0×). Given that this is a lightweight OLAP workload with many already-efficient queries, such a pattern is expected. The results confirm that CQO generally improves performance, but also underscore the importance of suppressing clarification when expected gains are low or baseline latency is minimal.

\begin{figure}[t]
  \centering
  \includegraphics[width=0.8\linewidth]{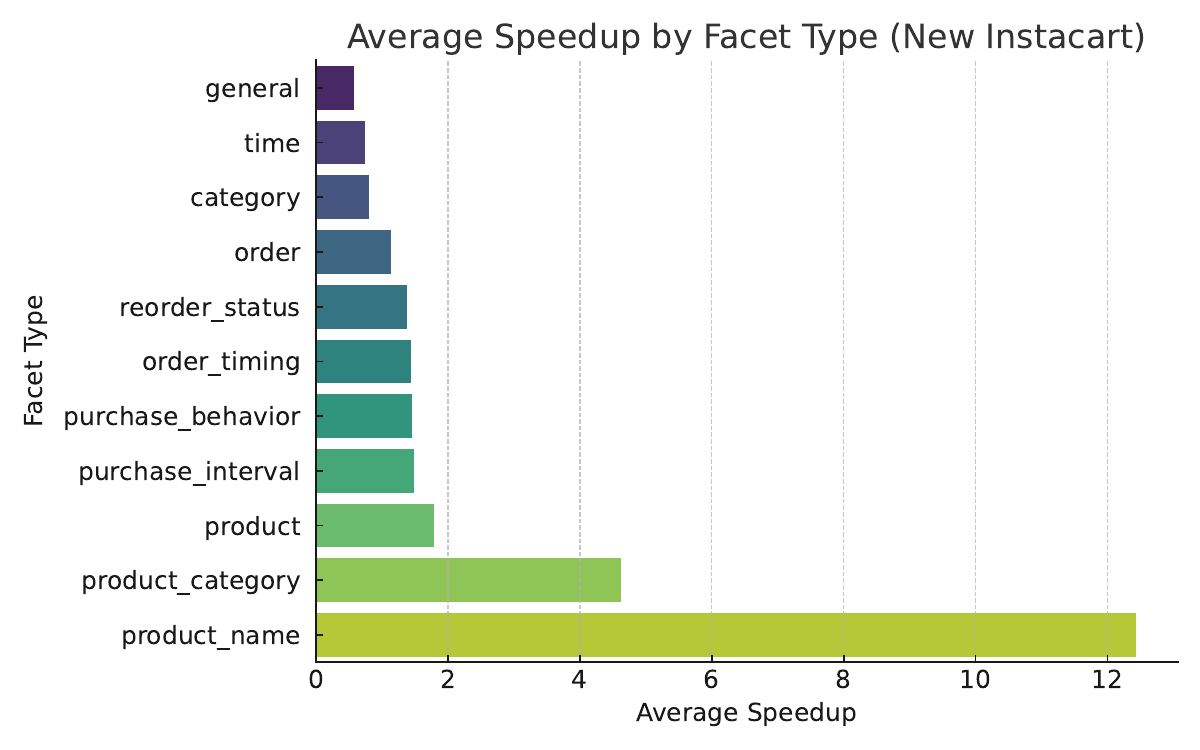}
  \caption{Average Speedup by Facet Type --- Instacart}
  \label{fig:Instacart_2}
\end{figure}

Figure~\ref{fig:Instacart_2} reveals that clarifications on facets like \texttt{Product Name}, \texttt{Product Category}, and \texttt{Purchase Time} yield the highest average speedups—often exceeding 1.4×—by enabling selective filters and index use. In contrast, vague or low-selectivity facets such as \texttt{User Attribute} or \texttt{General} provide little or negative benefit. This highlights the critical role of facet selection: effective clarifications reduce execution cost by adding meaningful constraints, while weak ones increase complexity without payoff.

\subsubsection{SciDocs dataset}

\begin{figure}[t]
  \centering
  \begin{subfigure}[t]{0.48\columnwidth}
  \centering
  \includegraphics[width=0.82\linewidth]{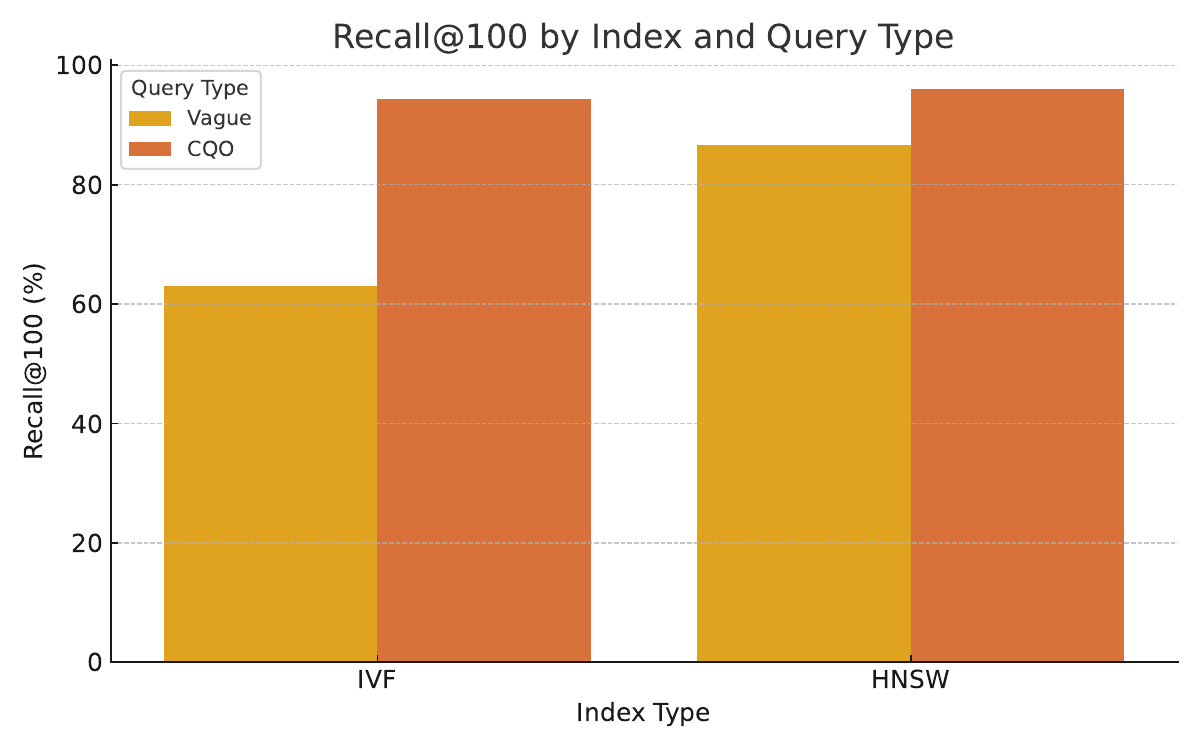}
  \caption{Recall@100 By Index And Query Type --- Scidocs}
  \label{fig:scidoc_recall}
  \end{subfigure}%
  \hfill
  \begin{subfigure}[t]{0.48\columnwidth}
  \centering
  \includegraphics[width=0.82\linewidth]{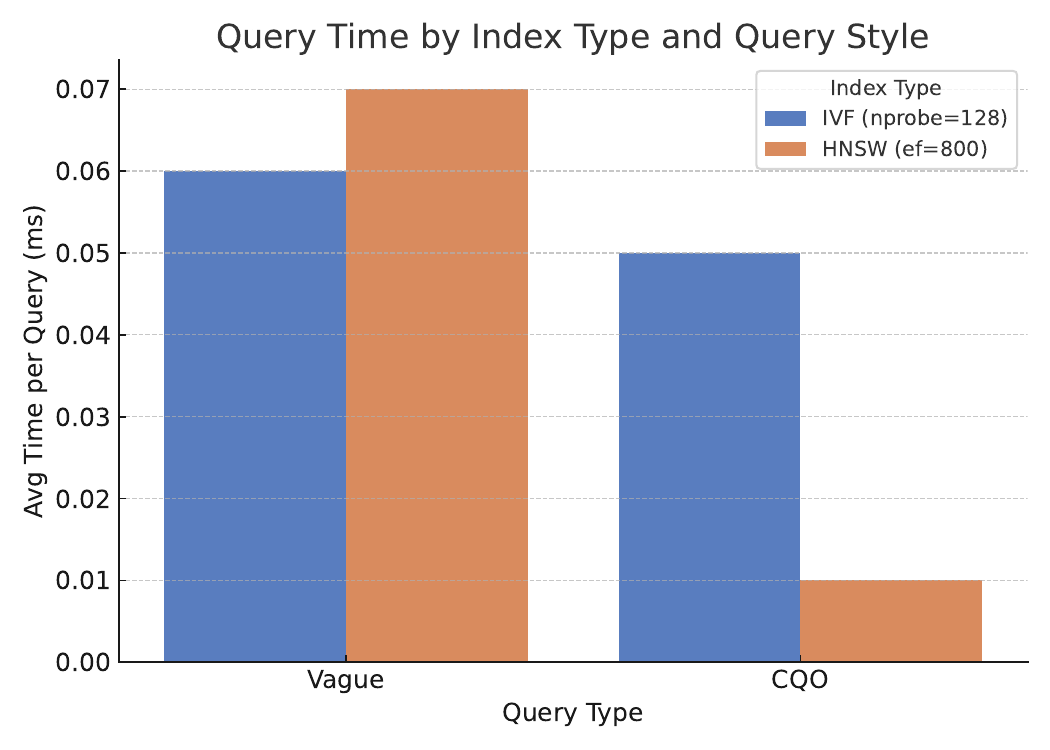}
  \caption{Query Time By Index Type And Query Style --- Scidocs}
  \label{fig:scidoc_Time}
  \end{subfigure}
  \caption{Recall vs.\ latency on the SciDocs dataset.}
  \label{fig:cqo-two-metrics}
\end{figure}

The detailed results are shown in Fig. \ref{fig:cqo-two-metrics}. The results show that CQO delivers consistent quality improvements across both index types. For IVF, Recall@100 increases from 63.0 under vague queries to 94.4 with CQO refinement. For HNSW, it improves from 86.6 to 96.0. In both cases, the clarified query helps narrow the semantic intent, aligning the vector more closely with the relevant region in the embedding space, leading to higher efficiency as well. Despite their architectural differences, both IVF’s cluster-based index and HNSW’s graph-based traversal benefit from this tighter alignment. The results suggest that CQO's lightweight clarifications can shift vague embeddings just enough to expose high-recall regions in the index, enhancing retrieval quality without requiring parameter tuning or increased compute.

\section{Conclusion}

\label{sec:conclusion}

In this paper, we propose Data-Aware Socratic Guidance (DASG), a framework that uses the Conversational Query Optimizer (CQO) and Maximum Information Utility (MIU) ranker to ask clarifying questions only when beneficial. By combining ambiguity detection and facet ranking across relational and vector queries, DASG improves efficiency and accuracy with low overhead. We are extending it to real human interactions and multi-round dialogue using MDP-based optimization.

\begin{acks}
\end{acks}

\bibliographystyle{unsrt} 
\bibliographystyle{ACM-Reference-Format}
\bibliography{main}

\begin{thebibliography}{10}

\bibitem{Li2024}
Boyan Li, Yuyu Luo, Chengliang Chai, Guoliang Li, and Nan Tang.
\newblock {The Dawn of Natural Language to SQL: Are We Fully Ready?}
\newblock {\em Proceedings of the VLDB Endowment}, 17(11):3318--3331, 2024.

\bibitem{Gao2024}
Dawei Gao, Haibin Wang, Yaliang Li, Xiuyu Sun, Yichen Qian, Bolin Ding, and Jingren Zhou.
\newblock {Text-to-SQL Empowered by Large Language Models: A Benchmark Evaluation}.
\newblock {\em Proceedings of the VLDB Endowment}, 17(5):1132--1145, 2024.

\bibitem{openai2024gpt4technicalreport}
OpenAI.
\newblock Gpt-4 technical report, 2024.

\bibitem{comanici2025gemini25pushingfrontier}
Google.
\newblock Gemini 2.5: Pushing the frontier with advanced reasoning, multimodality, long context, and next generation agentic capabilities, 2025.

\bibitem{hui2024qwen2}
Binyuan Hui, Jian Yang, Zeyu Cui, Jiaxi Yang, Dayiheng Liu, Lei Zhang, Tianyu Liu, Jiajun Zhang, Bowen Yu, Kai Dang, et~al.
\newblock Qwen2. 5-coder technical report.
\newblock {\em arXiv preprint arXiv:2409.12186}, 2024.

\bibitem{yu-etal-2018-spider}
Tao Yu, Rui Zhang, Kai Yang, Michihiro Yasunaga, Dongxu Wang, Zifan Li, James Ma, Irene Li, Qingning Yao, Shanelle Roman, Zilin Zhang, and Dragomir Radev.
\newblock {S}pider: A large-scale human-labeled dataset for complex and cross-domain semantic parsing and text-to-{SQL} task.
\newblock In {\em Proceedings of the 2018 Conference on Empirical Methods in Natural Language Processing}, Brussels, Belgium, October-November 2018. Association for Computational Linguistics.

\bibitem{li2024can}
Jinyang Li, Binyuan Hui, Ge~Qu, Jiaxi Yang, Binhua Li, Bowen Li, Bailin Wang, Bowen Qin, Ruiying Geng, Nan Huo, et~al.
\newblock Can llm already serve as a database interface? a big bench for large-scale database grounded text-to-sqls.
\newblock {\em Advances in Neural Information Processing Systems}, 36, 2024.

\bibitem{lei2024spider}
Fangyu Lei, Jixuan Chen, Yuxiao Ye, Ruisheng Cao, Dongchan Shin, Hongjin Su, Zhaoqing Suo, Hongcheng Gao, Wenjing Hu, Pengcheng Yin, et~al.
\newblock Spider 2.0: Evaluating language models on real-world enterprise text-to-sql workflows.
\newblock {\em arXiv preprint arXiv:2411.07763}, 2024.

\bibitem{gao2023texttosqlempoweredlargelanguage}
Dawei Gao, Haibin Wang, Yaliang Li, Xiuyu Sun, Yichen Qian, Bolin Ding, and Jingren Zhou.
\newblock Text-to-sql empowered by large language models: A benchmark evaluation, 2023.

\bibitem{Liu2020}
Xinyu Liu, Shuyu Shen, Boyan Li, Peixian Ma, Runzhi Jiang, Yuxin Zhang, Ju~Fan, Guoliang Li, Nan Tang, and Yuyu Luo.
\newblock {A Survey of Text-to-SQL in the Era of LLMs: Where are we, and where are we going?}
\newblock 18(9):1--20, jun 2025.

\bibitem{iyer2017learningneuralsemanticparser}
Srinivasan Iyer, Ioannis Konstas, Alvin Cheung, Jayant Krishnamurthy, and Luke Zettlemoyer.
\newblock Learning a neural semantic parser from user feedback, 2017.

\bibitem{Li2014}
Fei Li and Hosagrahar~V Jagadish.
\newblock {NaLIR}.
\newblock In {\em Proceedings of the 2014 ACM SIGMOD International Conference on Management of Data}, pages 709--712, New York, NY, USA, jun 2014. ACM.

\bibitem{yao-etal-2019-model}
Ziyu Yao, Yu~Su, Huan Sun, and Wen-tau Yih.
\newblock Model-based interactive semantic parsing: A unified framework and a text-to-{SQL} case study.
\newblock In {\em Proceedings of the Conference on Empirical Methods in Natural Language Processing and the 9th International Joint Conference on Natural Language Processing (EMNLP-IJCNLP)}, pages 5447--5458, Hong Kong, China, November 2019. Association for Computational Linguistics.

\bibitem{gur-etal-2018-dialsql}
Izzeddin Gur, Semih Yavuz, Yu~Su, and Xifeng Yan.
\newblock {D}ial{SQL}: Dialogue based structured query generation.
\newblock In {\em Proceedings of the 56th Annual Meeting of the Association for Computational Linguistics (Volume 1: Long Papers)}, pages 1339--1349, Melbourne, Australia, July 2018. Association for Computational Linguistics.

\bibitem{elgohary-etal-2021-nl}
Ahmed Elgohary, Christopher Meek, Matthew Richardson, Adam Fourney, Gonzalo Ramos, and Ahmed~Hassan Awadallah.
\newblock {NL}-{EDIT}: Correcting semantic parse errors through natural language interaction.
\newblock In {\em Proceedings of the Conference of the North American Chapter of the Association for Computational Linguistics: Human Language Technologies}, pages 5599--5610, Online, June 2021. Association for Computational Linguistics.

\bibitem{tian2024interactivetexttosqlgenerationeditable}
Yuan Tian, Zheng Zhang, Zheng Ning, Toby Jia-Jun Li, Jonathan~K. Kummerfeld, and Tianyi Zhang.
\newblock Interactive text-to-sql generation via editable step-by-step explanations, 2024.

\bibitem{qiu2025interactivetexttosqlexpectedinformation}
Luyu Qiu, Jianing Li, Chi Su, and Lei Chen.
\newblock Interactive text-to-sql via expected information gain for disambiguation, 2025.

\bibitem{neurdb-scis-24}
Beng~Chin Ooi, Shaofeng Cai, Gang Chen, Yanyan Shen, Kian-Lee Tan, Yuncheng Wu, Xiaokui Xiao, Naili Xing, Cong Yue, Lingze Zeng, Meihui Zhang, and Zhanhao Zhao.
\newblock Neurdb: An ai-powered autonomous data system.
\newblock {\em SCIENCE CHINA Information Sciences}, pages~--, 2024.

\bibitem{tablegpt}
Peng Li, Yeye He, Dror Yashar, Weiwei Cui, Song Ge, Haidong Zhang, Danielle Rifinski~Fainman, Dongmei Zhang, and Surajit Chaudhuri.
\newblock Table-gpt: Table fine-tuned gpt for diverse table tasks.
\newblock {\em Proc. ACM Manag. Data}, 2(3), May 2024.

\bibitem{Lu2025}
W.~Lu, J.~Zhang, J.~Fan, et~al.
\newblock Large language model for table processing: a survey.
\newblock {\em Frontiers of Computer Science}, 19:192350, 2025.

\end{thebibliography}

\appendix

\end{document}